\documentclass[prl,twocolumn,longbibliography]{revtex4}

\usepackage[utf8]{inputenc}
\usepackage{bm}
\usepackage{mathtools}
\usepackage{bbold}
\usepackage{amsfonts, amsmath, amsthm, amssymb}
\usepackage{hyperref}
\usepackage[normalem]{ulem}
\usepackage{color}
\usepackage{float}
\usepackage[dvipsnames]{xcolor}

\usepackage{ulem}

\begin{document}

\title{Symmetry: a fundamental resource for quantum coherence and metrology}
\author{Ir\'en\'ee Fr\'erot}
\email{irenee.frerot@lkb.upmc.fr}
\affiliation{Laboratoire Kastler Brossel, Sorbonne Universit\'e, CNRS, ENS-PSL Research University, Coll\`ege de France, 4 Place Jussieu, 75005 Paris, France}
\author{Tommaso Roscilde}
\email{tommaso.roscilde@ens-lyon.fr}
\affiliation{Univ Lyon, Ens de Lyon, CNRS, Laboratoire de Physique, F-69342 Lyon, France}
\date{\today}

\begin{abstract}
    We introduce a new paradigm for the preparation of deeply entangled states useful for quantum metrology. We show that when the quantum state is an eigenstate of an operator $A$, observables $G$ which are completely off-diagonal with respect to $A$ have purely quantum fluctuations, as quantified by the quantum Fisher information, namely $F_Q(G)=4\langle G^2 \rangle$. This property holds regardless of the purity of the quantum state, and it implies that off-diagonal  fluctuations represent a metrological resource for phase estimation. In particular, for many-body systems such as quantum spin ensembles or bosonic gases, the presence of { off-diagonal long-range order (for a spin observable, or for bosonic operators) directly translates into a metrological resource, provided that the system remains in a well-defined symmetry sector. The latter is defined {\it e.g.} by one component of the collective spin or by its parity in spin systems; and by a particle-number sector for bosons. Our results establish the optimal use for metrology of arbitrarily non-Gaussian quantum correlations in a large variety of many-body systems.}
\end{abstract}

\maketitle

\textit{Introduction.--} Many-body entanglement \cite{Horodeckietal2009} is a striking feature of composite quantum systems: it is responsible for the fundamental complexity of quantum mechanics; and it represents the main resource offered by quantum devices based on the coherent control of many-body systems. 
Among all currently envisioned quantum technologies, quantum metrology \cite{Pezzeetal2018,Degenetal2017} represents one of the most concrete applications of entangled states. 
The metrological task of interest to this work is phase estimation \cite{pezze2014quantum}, namely the reconstruction of the phase $\theta$ associated with a unitary transformation $U_{\theta} = e^{-i\theta G}$ acting on the state $\rho$ of the system, where $G$ is a Hermitian operator. Atomic clocks, gravimeters, magnetometers etc. are all based on this functioning principle. 
It is well established \cite{Pezzeetal2018} that systems of uncorrelated quantum particles do not give access to the ultimate precision on the phase $\theta$ allowed by quantum mechanics, but they are bounded by the so-called standard quantum limit of metrology. The ultimate precision (the so-called Heisenberg limit) can only be achieved in the presence of multipartite entanglement between the degrees of freedom \cite{hyllusetal2012,toth2012,Pezzeetal2018}. 

It is generally believed that entanglement useful for metrology is associated with states of high purity and small fluctuations of a specific observable, { whose paradigmatic example is offered by squeezed states} \cite{wineland1992,Ma2011PR,Pezzeetal2018}. Here, we unveil a different paradigm to produce arbitrarily mixed, yet highly valuable entangled states for metrology. The basic mechanism is the combination of two ingredients: { the states of interest are eigenstates of a symmetry operator; and at the same time they display strong correlations in observables which are completely off-diagonal in the eigenbasis of the symmetry operator.}

{ The sensitivity of a state to the unitary transformation $U_{\theta}$} is fundamentally related to its quantum Fisher information (QFI) $F_Q(G)$ \cite{BraunsteinC1994} associated with the generator $G$ of the unitary transformation, defined as 
\begin{equation}
F_Q(G) = 2 \sum_{nm} \frac{(p_n - p_m)^2}{p_n+p_m} |\langle n | G | m\rangle|^2~.
\label{e.QFI}
\end{equation}
Here $|n\rangle, |m\rangle$ are eigenstates of $\rho$ with eigenvalues $p_n, p_m$. The uncertainty $\delta \theta$ on the estimation of the phase $\theta$ allowed by the state $\rho$ is bounded by the QFI as per the quantum Cram\'er-Rao bound $(\delta \theta)^2\geq 1/F_Q(G)$. Hence a central task in quantum metrology is to prepare input states possessing a large QFI; and to identify observables $O$ whose $\theta$ dependence $\langle O \rangle_\theta = {\rm Tr} (U^\dagger_\theta O U_{\theta} \rho)$ can best reveal the sensitivity of the state to the unitary transformation of interest. This is expressed by the fact that the inequality $F_Q(G) \ge \xi_O^{-2}$ is tight, where $\xi_O^{-2}$ corresponds to the squared signal-to-noise ratio:
\begin{equation}
\xi_O^{-2} = \frac{|\langle[O,G ]\rangle_\theta|^2}{{\rm Var}(O)_{\theta}}~.
\label{e.xiO}
\end{equation}  
Here ${\rm Var}(O)_{\theta} = \langle O^2 \rangle_\theta - \langle O \rangle_{\theta}^2$, and we have used the fact that $i\partial_{\theta}\langle O \rangle_{\theta} = \langle [O, G] \rangle_\theta$. $O$ is an optimal observable for phase estimation when $\xi_O^{-2} = F_Q(G)$.

\begin{figure}[ht!]
\begin{center}
\includegraphics[width=\columnwidth]{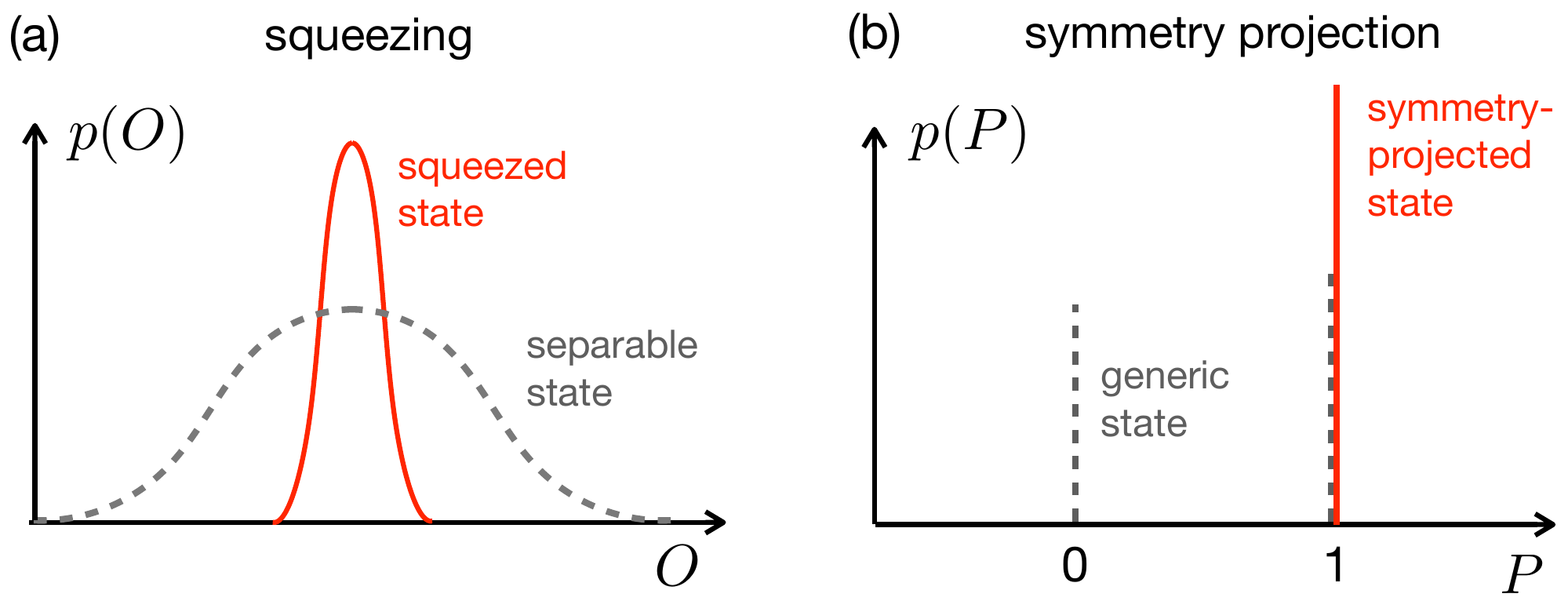}
\caption{ Different paradigms of metrological resources. (a) Squeezing of an observable $O$ with respect to that of separable states with large contrast (see text); (b) symmetry projection, with projector $P$. In both panels $p(A)$ ($A =O, P$) is the probability of finding a given eigenvalue of the operator $A$. }
\label{f.sketch}
\end{center}
\end{figure}

\textit{From squeezing to symmetry projection.--} An important paradigm for the use of entangled states to improve metrological sensitivity is offered by {\it squeezing} -- see Fig.~\ref{f.sketch}(a) for a cartoon. Squeezing is typically formulated for observables $O$ which are expressed as sums of local observables for a system of $N$ degrees of freedom, $O = \sum_{i=1}^{N} O_i$. 
Squeezing -- as associated with entanglement -- occurs when the parameter $\xi_O^{-2}$ is larger than that of {\it all} separable states, i.e. of mixtures of factorized states $\otimes_{i=1}^N |\psi_i\rangle$. For the latter states, the parameter $\xi_O^{-2}$ is typically maximized by maximizing the ``contrast'' $|\langle[O,G ]\rangle|$ -- { this is the case {\it e.g.} in ensembles of spin-1/2 particles \cite{Pezzeetal2018}. Since the contrast is generically not commuting with $O$, one has that for these states} ${\rm Var}(O) = \sum_i {\rm Var}(O_i) \sim N$. On the other hand, entangled squeezed states are associated with a reduction of ${\rm Var}(O)$ compared to that of separable states, while preserving a finite contrast. Entangled squeezed states of spin ensembles have been obtained in a large variety of platforms \cite{Esteve2008,Riedel2010,Muessel2014PRL,Bohnet2016,Bornet2023,Franke2023,Eckner2023}, and their first uses to increase the performance of {\it e.g.} atomic clocks have been recently demonstrated \cite{LouchetChauvet2010NJP,PedrozoPenafiel2020,Eckner2023}.

In this work, we propose a paradigm which is alternative to squeezing of the fluctuations of local observables. Our paradigm is { based on} {\it symmetry projection}: the observable $O$ of metrological interest becomes a {\it projector} $P$ onto a subspace of the Hilbert space, which can be generically associated with the eigenspace of a further observable $A$. { As a concrete example, for spin ensembles} one can define the collective-spin operator ${\bf J} = \sum_{i=1}^N {\bf S_i}$, where $\bf S_i$ is the $i$-th spin operator. $P$ can be thought of as the projector onto a subspace of well-defined magnetization along {\it e.g.} the $z$ axis, in which case $A =J^z$; or of well-defined parity of the magnetization, in which case $A = (-1)^{NS-J^z}$ (for $N$ spins of length $S$).
The states of our interest are states $\rho$ (pure {\it or} mixed) which live entirely in the sector projected by $P$, namely 
\begin{equation}
\rho = P \rho ~ (= \rho P = P \rho P)
\label{e.symmetry}
\end{equation} 
so that $\langle P \rangle = {\rm Tr}(\rho P) = 1$ and ${\rm Var}(P) = \langle P \rangle - \langle P \rangle^2 = 0$. Clearly these states are ``squeezed'' in terms of the fluctuations of $P$ compared with a generic state (see Fig.~\ref{f.sketch}(b)); we shall specify below under which circumstances they exhibit superior metrological properties compared to separable states. The difference with conventional squeezing is that $P$ can be a highly non-local operator -- this is clearly the case when $P$ is the projector on a parity sector. In the following we denote such states as symmetry projected.

In general, quantum states $\rho$ are especially sensitive to a transformation $U_{\theta}$ whose generator $G$ has a large variance in $\rho$, provided that the uncertainty is of quantum origin, namely that it stems from the fact that $[\rho,G]\neq 0$ { (as captured by the QFI)}. For symmetry-projected states, this condition is best realized when $G$ connects different symmetry sectors, namely when \begin{equation}
PGP = 0 ~~~~ PGQ + QGP = G
\label{e.G}
\end{equation}
where $Q=\mathbb{1}-P$. In this case $G$ is completely off-diagonal on the eigenbasis of $\rho$. Our main result is that, in this case, all fluctuations of $G$ in $\rho$ are of quantum-mechanical origin; that they are metrologically useful, regardless of the purity of $\rho$, that is: $4{\rm Var}(G) = F_Q(G)$; and that the sensitivity of the projector to the unitary transformation, $\xi_P^{-2}$ (defined as in Eq.~\eqref{e.xiO}) equals $F_Q(G)$. Our result generalizes and strengthens a similar theorem reported in Ref.~\cite{Nolan2017} in the context of spin systems, in which the classical Fisher information of the magnetization distribution in parity eigenstates is shown to coincide with the QFI.

Our results show that symmetry-projected states with strong off-diagonal correlations (namely with a large ${\rm Var}(G)$, ideally scaling as $\sim N^2$ if $G$ is the sum of local quantities) are highly entangled; that this holds true regardless of their purity; and that they represent fundamental resources for quantum metrology. Below, we shall offer examples for the relevance of such states in the context of Bose-Einstein condensates and of quantum spin ensembles. 
Many recent experiments on quantum many-body devices have demonstrated the capability of preparing correlated states \cite{Richerme2014,Jurcevic2014,Alaouietal2022,chen2023_2,chen2023}. Symmetry preservation can then act as a very effective tool to certify which experiments prepare highly entangled states, as opposed to states with classical correlations \cite{Frerotetal2023}.

\textit{Symmetry-projection theorem.--} We begin the discussion of our results with the following 

\noindent\textit{Theorem.} Consider a projector $P$ on a subspace, and a state $\rho$ having support only in that subspace -- i.e. fulfilling Eq.~\eqref{e.symmetry}. Consider then a unitary transformation $U_\theta=e^{-i\theta G}$ { whose generator $G$ sends} vectors in the support of $P$ onto the orthogonal subspace: $PGP=0$. Then 
    \begin{equation}
        \langle P \rangle_\theta = {\rm Tr}(U_{\theta}^\dagger P U_\theta \rho ) = 1 - \theta^2 \langle G^2 \rangle + {\cal O}(\theta^3) ~,
        \label{eq_Ptheta}
    \end{equation}
 where $\langle ... \rangle = {\rm Tr}(\rho ...)$.  As a consequence:
\begin{equation}
   \xi^{-2}_P = 4 \langle G^2 \rangle = F_{\rm Q}(G)
   \label{eq_equality_chain}
\end{equation}
where $\xi_P^{-2}$ is defined as in Eq.~\eqref{e.xiO} in the limit $\theta \to 0$.

\begin{proof}
We Tailor-expand $P_\theta = U_\theta^\dagger P U_\theta =   P + i\theta(GP - PG) + \theta^2 G P G - (\theta^2 / 2)(G^2 P + P G^2) + {\cal O}(\theta^3)$, and evaluate $\langle P \rangle_\theta = {\rm Tr}[P_\theta \rho]$. Since $P\rho = \rho = \rho P$, using circular invariance of the trace we have that ${\rm Tr} [\rho(GP - PG)] = {\rm Tr}[G(P\rho - \rho P)] = 0$. Similarly, we have ${\rm Tr}(\rho P G^2) = {\rm Tr}(G^2 P \rho) = \langle G^2 \rangle$. Then ${\rm Tr}( G P G \rho)={\rm Tr}(G P G P \rho) = 0$ where we used $P\rho=\rho$ and $PGP=0$. Hence Eq.~\eqref{eq_Ptheta} follows. From this, we find $[\partial_\theta \langle P(\theta) \rangle]^2 = 4\theta^2 \langle G^2 \rangle^2 + {\cal O}(\theta^3)$. Since $P^2 = P$, we also have that ${\rm Var}[P(\theta)] = \langle P \rangle - \langle P \rangle^2 = \theta^2 \langle G^2 \rangle + {\cal O}(\theta^3)$, from which $\xi_P^{-2}=4\langle G^2 \rangle$. Since for an arbitrary observable $O$ we have $\xi_O^{-2} \le F_{\rm Q}(G) \le 4 {\rm Var} (G) \le  4 \langle G^2 \rangle$ \cite{pezze2014quantum,Pezzeetal2018}, we conclude that for $O=P$ this chain of inequalities becomes a chain of equalities, and hence Eq.~\eqref{eq_equality_chain} follows.
\end{proof}

We emphasize that in general  $4 \langle G^2 \rangle \geq F_{\rm Q}(G)$, while  the equality $4 \langle G^2 \rangle = F_{\rm Q}(G)$ is always realized by {\it pure} states  $\rho = |\psi\rangle \langle \psi | $ with $\langle G \rangle = 0$. Finding mixed states that satisfy this equality is instead highly non-trivial. From a physical point of view, the equality implies that \emph{all} fluctuations of $G$ in state $\rho$ have a quantum-coherent origin, as they stem from $G$ being completely off-diagonal in the basis of $\rho$, and this independently of the purity of $\rho$. In turn, all fluctuations of $G$ have an immediate metrological relevance, as they translate into the sensitivity of the average projector $\langle P \rangle$ to the $U_{\theta}$ transformation. Moreover our result has immediate implications for entanglement certification \cite{Frerotetal2023}. Indeed, if $G$ is a sum of local observables $G = \sum_i G_i$, the QFI serves as a witness for multipartite entanglement, excluding separability of the state when $F_Q(G) > 4 \sum_i {\rm Var}(G_i)$ \cite{Gessneretal2017}. In the case of symmetry-projected states, if one further assumes that $PG_i P = 0$ for all $i$, this criterion simply becomes $\langle G^2 \rangle > \sum_i \langle G_i^2\rangle$, which is automatically verified in the presence of positive correlations $\langle G_i G_j \rangle > 0$ between degrees of freedom $i\neq j$. Hence symmetry projection translates positive correlations for off-diagonal local observables into an entanglement witness.

 
 The remaining question is then the following: where can one find symmetry-projected states which have significant off-diagonal correlations, implying strong multipartite and metrologically useful entanglement? In the next sections, we show that symmetry projection can be enforced in at least three ways: relying on super-selection rules; relying on the symmetry of Hamiltonians; or projecting the state of the system on a specific symmetry sector by non-destructive post-selection. We will illustrate these various possibilities in the case of Bose-Einstein condensates in a system of coupled bosonic modes; and in the case of interacting quantum spin (or qubits).  
 

\begin{center}
\begin{figure*}[ht!]
\includegraphics[width=\textwidth]{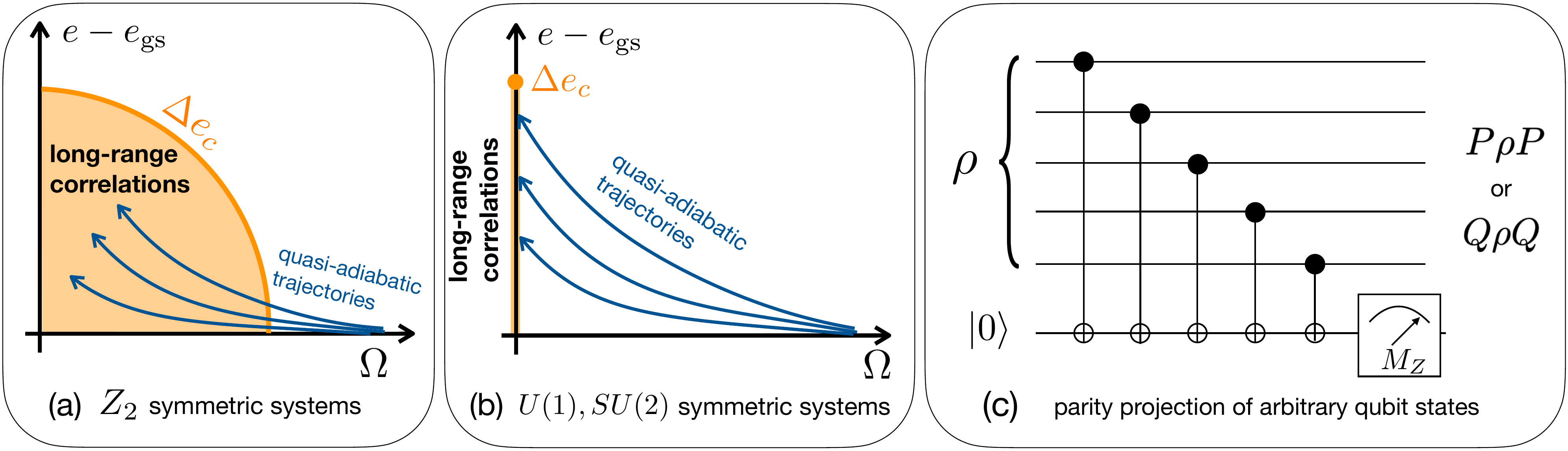}
\caption{Preparation of long-range correlated quantum states in a fixed parity sector. (a,b) Preparing an initial state polarized along the transverse field $\Omega$, and then ramping down $\Omega$ towards low-energy states of the parity-preserving Hamiltonian, allows one to prepare resource-states for metrology even at finite energy density and extensive entropy. The critical energy density difference $\Delta e_c$ with respect to the ground-state energy density $e_{\rm gs}$ marks the loss of long-range order. (c) Quantum-circuit projection of an arbitrary qubit state $\rho$ onto a well-defined parity sector, depending on the measurement  $M_Z$ of $\sigma^z$ for the ancilla qubit.}
\label{f.correlations}
\end{figure*}
\end{center}

\textit{Heisenberg-scaling sensitivity of tunneling rates with Bose-Einstein condensates (BEC).--}
We consider the metrological task of estimating the tunneling rate $J$ between { two sets of bosonic modes $A$ and $B$. The tunneling process is described by the Hamiltonian $H=J G$ where $G = -\frac{1}{2}\sum_{i=1}^N (a_i^\dagger b_i + b^\dagger_i a_i)$, with $a_i$ (resp. $b_i$) describing the bosonic modes in subsystem $A$ (resp. $B$); we imagine the modes to be pairwise coupled, as {\it e.g.} in two coupled chains or two coupled layers of localized modes.  If one considers $N$ independent bosons, each localized a mode pair $(a_i, b_i)$, the quadratic error} on the estimation of $J$ scales as $1/N$ (standard quantum limit). { On the other hand, we can show that two BECs, one delocalized over the $\{a_i\}$ modes and the other over the $\{b_i\}$ modes, allow for an estimation of $J$ with a quadratic error scaling as $1/N^2$ (Heisenberg scaling).}

In the framework of our theorem, we consider an initial state of the form $\rho_A \otimes \rho_B$ with the two sets of modes $A$ and $B$ having a fixed boson number $N_A=\sum_{i=1}^N a_i^\dagger a_i$ (resp. $N_B = \sum_{i=1}^N b_i^\dagger b_i$): $P_{N_A} \rho_A = \rho_A$, with $P_{N_A}$ the projector onto the states with fixed boson number $N_A$ (as well as $N_B$, imagining to have a fixed total boson number). The generator $G$ is purely off-diagonal with respect to the boson number in $A$ (or $B$). Coupling the two sets of modes exposes the state to the unitary transformation $e^{-itJG}$. According to our theorem, the QFI for the estimation of $tJ$ (and hence of $J$) is given by $4\langle G^2 \rangle$, and it is saturated by measuring the time evolution of $\langle P_{N_A} \rangle$, {\it i.e.} the probability to find $N_A$ bosons in the $A$ system (or equivalently, by measuring $P_{N_B}$ in the $B$ system). One finds:
\begin{equation}
    F_Q(G) = 4 \langle G^2 \rangle = N_A + N_B + \sum_{i,j}(\langle a_i^\dagger a_j \rangle \langle b_j^\dagger b_i \rangle + {\rm h.c.})   
    \label{eq_FQ_bosons}
\end{equation}
In the limit of ideal BECs with perfect off-diagonal long-range order, one has $\langle a_i^\dagger a_j \rangle = N_A / N := n_A$ and $\langle b_i^\dagger b_j \rangle = N_B / N := n_B$, hence $4\langle G^2 \rangle = N(n_A + n_B) + 2N^2 n_A n_B$.  { The Heisenberg scaling $F_Q(G) \sim N^2$ is fundamentally related to the entanglement between the $N$ modes of $A$ (resp. $B$) associated with the existence of a BEC}. Indeed, for a fully separable state $\rho_A = \sum_\lambda p_\lambda \otimes_{i=1}^N \rho_i(\lambda)$ with $p_\lambda$ an { arbitrary} probability distribution and $\rho_i(\lambda)$ a density matrix for mode $a_i$ (and similarly for the $B$ system), one has the inequality \cite{Gessneretal2017} $F_Q(G) \le \sum_i \langle (a_i^\dagger b_i + b_i^\dagger a_i)^2 \rangle \sim N$. More explicitly, taking uniform densities $\langle a_i^\dagger a_i \rangle = N_A / N = n_A$ (and similarly for $B$), one finds the separable bound $F_Q(G) \le F_{\rm sep} = N (n_A + n_B + 2 n_A n_B)$, leading to a quadratic error on the estimation of $J$ scaling no faster than $1/N$. { A uniform ideal BEC on the $B$ system (namely $\langle b_i^\dagger b_j \rangle = N_B / N$) leads to a QFI in the form $F_Q(G) = F_{\rm sep} + (N_B/N)\sum_{i\neq j} \left ( \langle a_i^\dagger a_j \rangle + {\rm c.c.} \right )$. 
Therefore, the condition $\sum_{i \neq j} {\rm Re} \langle a_i^\dagger a_j \rangle > 0$ witnesses mode entanglement, since it violates the bound for separable states. This illustrates the fact that entanglement generated solely by the indistinguishability of quantum particles represents a consistent resource for metrology \cite{Morrisetal2020}. In the case of atomic BECs, the symmetry of the initial state $\rho$ corresponds to a super-selection rule enforced by atom-number conservation. Our metrological protocol hence requires only post-selection on specific initial numbers $N_A, N_B$ of atoms in the two modes.  }

\textit{Quantum spin ensembles.--} 
We now consider systems of $N$ quantum $S=1/2$ spins 
\footnote{Although most of our results can be readily generalized to arbitrary spins $S$.}, prepared initially in a { coherent spin state} $|\psi_0\rangle=|\rightarrow_x\rangle^{\otimes N}$ fully polarized along the $x$ axis, and where the parity projector $P_x = (1+\prod_{i=1}^N \sigma_i^x)/2$ is both fixed by the initial state, $P_x|\psi_0\rangle=|\psi_0\rangle$, and preserved during the dynamics. In particular, $P_x$ is conserved by all interaction Hamiltonians of the form $H_{\rm int}= - \sum_{i,j} \sum_{a \in \{x,y,z\}} J_{ij}^a \sigma_i^a \sigma_j^a$, namely XYZ models with arbitrary two-body interactions, as well as by Hamiltonians describing an external magnetic field oriented along $x$: $H_{\rm field}=- \sum_i \Omega_i \sigma_i^x$. Our theorem then guarantees that, at all times, spin fluctuations for components in the $yz$ plane are fully quantum-coherent; and that the { sensitivity of the state to spin rotations around an axis in the same plane} is optimally exploited by simply measuring the parity projector $P_x$ itself. A simple symmetry constraint entails therefore very strong consequences on the nature of fluctuations in the system, and their potential application for quantum metrology. In particular, our result applies to { widely different model Hamiltonians for quantum simulators whose interactions possess different symmetries but all preserve the parity $P_x$}, such as transverse-field Ising models $H_{\rm Ising}=-\sum_{i,j} J_{ij} \sigma_i^z \sigma_j^z  -\Omega \sum_i \sigma_i^x$ (with $Z_2$ symmetry) \cite{Monroe2021RMP, Scholl2021,Ebadi2021,King2022}; XY models $H_{\rm XY} = -\sum_{i,j} J_{ij} (\sigma_i^x \sigma_j^x + \sigma_i^y \sigma_j^y)$ (with $U(1)$ symmetry) \cite{BrowaeysL2020,Monroe2021RMP}; and XXZ (Heisenberg) models $H_{\rm XXZ}=-\sum_{i,j} J_{ij}(\sigma_i^x \sigma_j^x + \sigma_i^y \sigma_j^y - \sigma_i^z \sigma_j^z)$ (with $SU(2)$ symmetry) \cite{Mazurenko2017,Jepsen2020,Sun2021}. Notice that the one-axis-twisting Hamiltonian (OAT), whose dynamics is implemented in many recent experiments \cite{Riedel2010,Bohnet2016,Song2019,Bornet2023,Franke2023},  may be viewed as a special instance of XY models, by writing $H_{\rm OAT} = \chi J_z^2 /N = \frac{\chi}{N} \left[ {\bf J}^2 - (1/4)\sum_{i,j} (\sigma_i^x \sigma_j^x + \sigma_i^y \sigma_j^y) \right ]$, where the conserved total spin ${\bf J}^2$  is just a constant offset. Therefore our results provide the optimal strategy to exploit metrologically all the non-Gaussian spin states \cite{Strobeletal2014, Pezzeetal2018, Comparinetal2022} produced by the OAT dynamics after the appearance of maximal spin squeezing (for a time $t\sim N^{1/3}$) and before the appearance of the Schr\"odinger cat state (for a time $t = \pi N/(2\chi)$), by measuring the sensitivity of the parity $P_x$ to rotations around the $y$ axis.  

The initial state is the ground state of $H_{\rm field}$. A general strategy to prepare metrologically useful states is then to evolve $|\psi_0\rangle$ with any of the parity-preserving Hamiltonians above $H=H_{\rm int} + H_{\rm field}$, including possibly time-dependent parameters, towards a state displaying long-range correlations for spin components in $yz$ plane. Such long-range correlations typically emerge when the system displays an ordered phase at low energy, spontaneously breaking the $Z_2$, $U(1)$ or $SU(2)$ symmetry of the Hamiltonian $H_{\rm int}$ in the thermodynamic limit. Maximal correlations are expected in the ground state of $H_{\rm int}$, which requires to implement an adiabatic evolution starting from the initial state, namely the ground state of $H_{\rm field}$. In a realistic (quasi-adiabatic) evolution implemented in experiments  in which \emph{e.g.} $H_{\rm field}$ is slowly ramped down  (see Fig.~\ref{f.correlations}(a-b)), the final state has finite energy density above the ground state. One may then na\"ively conclude that the resulting state is akin to a thermal mixed state,  implying that quantum correlations have acquired a short-ranged natureeven when total correlations are long-ranged \cite{MalpettiR2016,Frerotetal2022,KuwaharaS2022,Scheieetal2024}.
Yet, at variance with this thermal picture of correlations, our theorem establishes that the preparation of correlated state corresponds to the onset of fully quantum correlations for off-diagonal observables, regardless of the purity of the final state, whenever parity is preserved.
Typical phase diagrams in the external-field/energy-density plane are illustrated in Fig.~\ref{f.correlations}(a,b): whenever parity-preserving quasi-adiabatic evolutions end up in a {\it long-range} ordered phase, they give rise to macroscopic quantum Fisher information (i.e scaling as $N^2$) for a completely off-diagonal collective-spin component, such as the $y$ or $z$ component of the collective spin (uniform or staggered). For instance, ferromagnetic transverse-field Ising models develop $F_Q(J^z) \sim O(N^2)$ at finite energy density when the connectivity of their interactions is sufficiently high ({\it e.g.} in dimensions $d \geq 2$ with nearest-neighbor interactions). Similarly, $F_Q(J^y) \sim O(N^2)$ at finite energy density in ferromagnetic XY models when $J_{ij} \sim 1/r_{ij}^{\alpha}$ with $r_{ij}$ the distance between sites, and $\alpha < d+2$ in $d\leq 2$ dimensions \cite{Bruno2001}; and for arbitrarily short-ranged interactions in $d=3$ -- the latter results holds as well for the SU(2) symmetric Heisenberg model.

\textit{Parity projection.--} An alternative route towards preparing highly entangled states is offered by projecting an arbitrary qubit state $\rho$ onto a given parity sector. Active parity projection has the power of turning even a classically correlated state into an entangled one. 
Parity projection can be achieved with a conceptually simple quantum circuit, namely a sequence of C-NOT gates with an auxiliary qubit, followed by the measurement of the latter. This is illustrated in Fig.~\ref{f.correlations}(c) for parity projection in the $z$ basis: starting from an arbitrary basis state $|\vec z\rangle =|z_1, \dots, z_N\rangle$ with $z_i\in\{0, 1\}$, the sequence of C-NOT gates realizes the linear transformation $|\vec z\rangle \otimes |0\rangle \to |\vec z \rangle \otimes |(1+P(\vec z))/2\rangle$ with $P(\vec z)$ the parity of the magnetization of $|\vec z\rangle$.
With an arbitrary input state $\rho \otimes |0\rangle \langle 0|$, measuring the auxiliary qubit then allows one to implement the parity projection $\rho \to \rho_{\rm even} = P\rho P$ (resp. $\rho_{\rm odd} = Q\rho Q$) by post-selecting the $0$ (resp. 1) measurement outcome.
Most importantly, the transformation $\rho \to P \rho P + Q \rho Q$ preserves two-body correlations for the $x$ spin components, namely: ${\rm Tr}(\rho \sigma_i^x \sigma_j^x) = {\rm Tr}[(P \rho P + Q \rho Q)\sigma_i^x \sigma_j^x]$ for all $i,j$.
Indeed $\sigma_i^x \sigma_j^x = P\sigma_i^x \sigma_j^x P + Q \sigma_i^x \sigma_j^x Q$, since $\sigma_i^x \sigma_j^x$ is parity preserving. Hence post-selecting the state with the strongest correlations between $P\rho P$ or $Q\rho Q$ allows one to turn the arbitrary correlations of $\rho$ into quantum ones.
The simplest example of this construction is obtained by taking $\rho$ as the coherent spin state $|\rightarrow_x\rangle^{\otimes N}$: the post-selected even / odd parity states are cat states $(|\rightarrow_x\rangle^{\otimes N} \pm |\leftarrow_x\rangle^{\otimes N})/\sqrt{2}$, which have maximal QFI for rotations around the $x$ axis, $F_Q(J_x)=N^2$.\\

\textit{Conclusions.--} 
We have shown that whenever the quantum dynamics { of many-body systems} remains constrained to a given symmetry sector, all correlations of off-diagonal observables maintain a fully quantum nature, regardless of the purity of the quantum state; and that they directly translate into a metrological resource. Our result opens the path to the systematic certification and metrological use of a { broad class of entangled states}, such as {\it e.g.} non-Gaussian entangled spin states which are naturally prepared in quantum spin ensembles, going beyond the case of spin-squeezed states.

\textit{Acknowledgements.}-- TR acknowledges support of PEPR-Q ``QubitAF".

\bibliography{biblio}

\end{document}